\documentclass[draftcls, onecolumn]{IEEEtran}
\usepackage{amsfonts}
\usepackage{amssymb}
\usepackage{graphicx}
\usepackage{amsmath}
\renewcommand{\baselinestretch}{1.0}
\begin{document}

\title{Model and performance evaluation of field-effect transistors based on epitaxial graphene on SiC }
\author{Martina Cheli, Paolo Michetti and Giuseppe Iannaccone\\
Dipartimento di Ingegneria dell'Informazione: Elettronica, Informatica, Telecomunicazioni, via Caruso 16, 56100 Pisa, Italy\\
email: {\tt \{martina.cheli, paolo.michetti, g.iannaccone\}@iet.unipi.it}}

\maketitle

\bibliographystyle{IEEEtran}

\begin{abstract}
In view of the appreciable semiconducting gap of $0.26$~eV observed in  
recent experiments,
epitaxial graphene on a SiC substrate seems a promising channel material
for FETs. Indeed, it is  two-dimensional - and therefore does not  
require prohibitive lithography -
and exhibits a wider gap than other alternative options, such as  
bilayer graphene.
Here we propose a model and assess the achievable performance of a  
nanoscale FET based on epitaxial
graphene on SiC, conducting an exploration of the design parameter  
space.
We show that the current can be modulated by 4 orders of magnitude; 
for digital applications  an $I_{\rm on}/I_{\rm off}$ ratio of $50$  
and a subthreshold slope of $145$~mV/decade can be obtained with a supply voltage of $0.25$~V. This  
represents a significant progress towards solid-state integration of  
graphene electronics, but not
yet sufficient for digital applications.

\end{abstract}

\section{Introduction}
Epitaxially grown graphene on SiC provide potential  for large scale integration of graphene electronics. 
The first challenge to the use of graphene as a channel material for FETs is to induce a reasonable gap for room temperature operation. Recently Zhou {\it et al.}~\cite{Zhou2007} have experimentally demonstrated that a graphene layer, epitaxially grown on a SiC substrate, can exhibit a gap of about $0.26$~eV, measured by angle-resolved photo-emission spectroscopy. 
The gap is probably due to symmetry breaking between the two sublattices forming the graphene crystalline structure, as also confirmed by recent density functional calculations~\cite{Peng2008,kim2008}.
According to the authors of Ref.~\cite{Zhou2007}, this method of inducing a gap is very easy and reproducible; in addition, the thickness of graphite grown on SiC can be precisely controlled to be either single- or multiply layered depending on growth parameters~\cite{Brar2007}.
From a manufacturability point of view it is also extremely promising, since it would be highly convenient to prepare an entire substrate of graphene on an insulator and then obtain single device and integrated circuit through patterning~\cite{Kedzierski2008}.
\\From its isolation~\cite{Geim2004,Novoselov2005}, graphene has attracted the attention of the
scientific community due to its exceptional physical properties, such as an electron mobility exceeding more than $10$ times that of silicon wafers~\cite{Geim2007}, and in view of its possible
applications in transistors~\cite{Dai2008} and in sensors~\cite{Schedin2007}.
To induce a gap in graphene structures, several methods have been used:  lateral confinement in graphene ribbons ~\cite{Dai2008,son2006}  carbon nanotubes~\cite{martel1998}, impurity doping~\cite{Biel2009}, or a combination of single and bilayer graphene regions~\cite{Ohta2006,Fiori2008,Wu2008}. Unfortunately, they all face different problems.
\\Carbon nanotubes exhibit large intrinsic contact resistance and are difficult to pattern in a reproducible way; the inability to control tube chirality, and thus whether or not they are metallic or semiconducting, make solid state integration still prohibitive.
Graphene nanoribbons ~\cite{Dai2008,son2006} allow to obtain a very interesting 
device behavior~\cite{Fiori2007}, but require extremely narrow ribbons with
single-atom precision, since a difference of only one dimer line in the width may yield a quasi-zero gap nanoribbon.
Bilayer graphene exhibits a gap in the presence of a perpendicular
electric field, but the range of applicable bias can only induce a gap of $100-150$~meV,  not sufficient to
obtain a satisfactory behavior in terms of $I_{\rm on}/I_{\rm off}$ ratio~\cite{Fiori2008}.
Graphene on SiC can is a two-dimensional material, thus does not require extremely sophisticated
lithography, and provides a higher energy gap: for the sake of comparison, a $0.26$~eV energy 
gap would require an  armchair nanoribbon of width smaller than $3$~nm, or nanotubes 
with diameter smaller than $2$~nm.
\\In this work we present a semi-analytical model of an FET
with a channel of epitaxial graphene grown on a SiC substrate, 
where the band structure, the electrostatics, thermionic and band-to-band tunneling
currents are carefully accounted for. 
On the basis of our model, we assess the achievable device performance
through an exploration of the device parameter space, and gain
understanding of the main aspects affecting device operation.

\section{ Model}
We adopt the Tight Binding (TB) Hamiltonian for single layer graphene on
SiC that was proposed by Zhou {\it et al.}~\cite{Zhou2007}.
The empirical TB valence $(-)$ and conduction $(+)$ bands of a single epitaxial layer of graphene on SiC, read:
\begin{eqnarray}
 E_\pm(k_x,k_y)=\pm \sqrt{m^2+t^2|f(\mathbf{k})|^2},
\end{eqnarray}
where $t$ is the in-plane hopping term ($2.7$~eV), $m=0.13$~eV
is an empirical potential energy shift between the two inequivalent graphene sublattices due to interaction with the SiC substrate,  and $f(\mathbf{k})$ is
the off-diagonal element of the considered
Hamiltonian~\cite{Zhou2007}.
In the six Dirac points of the graphene Brillouin zone, where
$f(\mathbf{k})$ is zero, there is a finite energy gap $E_g=2m$, corresponding to the channel conduction
minimum $E_{CC} = m -q\phi_{ch}$ and the channel valence maximum $E_{VC} = -m -q\phi_{ch}$, where $q$ is the electron charge and $\phi_{ch}$ the
self-consistent potential in the central region of the channel.

\begin{figure}[!ht]
\centering
\includegraphics[width=7.0cm, height=12cm]{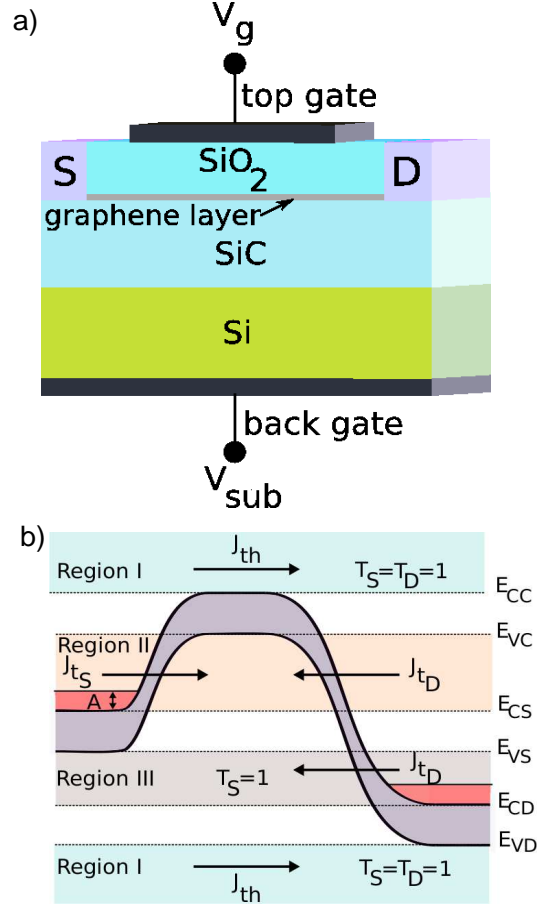}
\caption{a) Schematic picture of a graphene on SiC transistor. 
  The grey  line between SiO$_2$ and SiC oxide represents the graphene
  plane acting as device channel. Source (S) and drain (D) contacts are also in graphene. b) Profile band structure along the transport direction. 
  The dashed lines mark the energy region in which it is possible to
  have thermionic current (J$_{th}$), 
  tunneling current from source to channel (J$_{tD}$) and tunneling
  current from drain to channel (J$_{tS}$).
}
\label{device}
\normalsize
\end{figure}

The device under consideration, depicted in ~\ref{device}(a), is a transistor
with a channel of epitaxial graphene on a SiC substrate of thickness
$t_{sub}=100$~nm, with a top gate separated by a SiO$_2$ layer of thickness $t_{ox}$. 
In~\ref{device}(b) we have sketched the band edge profiles along the
transport direction $\hat{x}$, where $E_{Ci}$ and $E_{Vi}$
respectively represent the conduction and valence band edges in the
three different regions denoted by $i$=$S$, $D$, $C$ (Source, Drain, Channel). 
Source and drain contacts are $n^+$ doped, with molar fraction
$\alpha_{D}$, which translates into an energy difference $A$ between
the electrochemical potential $\mu_S$ ($\mu_D$) and the conduction
band edge $E_{CS}$ ($E_{CD}$) at the source (drain) contact.
The potential is set to zero at the source and to $V_{ds}$ at the drain
contact. In the center of the channel $\phi_{ch}$  is imposed by  vertical electrostatics.
We assume, as usual,  complete phase randomization
along the channel, which is particularly important 
because it allows us to neglect the effect of resonances in the presence
of tunneling barriers.

Exploiting the Gauss theorem we can write the  surface
charge density in the central part of the channel as   
\begin{eqnarray}
  Q=-C_{g}\left(V_{g}-V_{FBt}-\phi_{ch} \right)-C_{sub}\left(V_{sub}-V_{FBb}-\phi_{ch} \right),
  \label{eq:electrostatics}
\end{eqnarray}
where $C_{g}=\epsilon_{SiO_2}/t_{ox}$ ($C_{sub}=\epsilon_{SiC}/t_{sub}$) is the capacitance 
per unit area between the channel and the top gate (back gate),
$V_{g}$ ($V_{sub}$) is the top gate (back gate) voltage, $V_{FBt}$ ($V_{FBb}$) is the
flat-band voltage of the top gate (back gate), which we set to $-0.4$~eV.

The  transit time of the device in the channel has been estimated as  $\tau_t=\frac{Q_{th}L_{C}}{J_{th}}\approx 10^{-16}$~s 
 where $Q_{th}$ and
$J_{th}$ are the thermionic charge and current, respectively, 
 $L_{C}=20$~nm is the channel length. 
 
In certain spectral regions, for example in the valence band when the device is in the off state, 
carriers are quasi confined by tunneling barriers, and  can dwell in the channel for a much longer time 
and be subject to some degree of inelastic relaxation, even if transport in the conduction band is practically
ballistic.
To consider this effect, we have therefore included a degree of inelastic scattering that 
leads to energy relaxation.

In steady-state conditions, considering an infinitesimal element of area $dk_xdk_y$ in the wave-vectors space, charge distribution in the channel is obtained as a balance between two types of charge exchange processes with the contacts: one elastic, and one inelastic.
\\We can write the electron charge in the channel as the sum of two  contributions: $N_+^e$ and $N_-^e$.
$N_+^e=f_+^eL_Cn_{2D}$  ($N_-^e=f_-^eL_Cn_{2D}$) represents the density of forward (backward) going electrons,  
$f_{\pm}^e\left(k_x, k_y \right)$ denotes the occupation factors of forward $(+)$ and backward $(-)$  states
 in the channel and $n_{2D}\left(k_x, k_y \right)$ is the 2-dimensional density of states in the $k$-space.
For each contribution we can write a rate equation in steady-state conditions:
\begin{eqnarray}
\frac{dN_+^e}{dt}=J_S^+- J_D^++\left( 1-T_S\right)f_-^ev_x-\left( 1-T_D\right)f_+^ev_x +\frac{f_S^e-f_+^e}{\tau_S}L_Cn_{2D}  =0
\label{Npiu}
\end{eqnarray} 
\begin{eqnarray}
\frac{dN_-^e}{dt}=-J_S^-+ J_D^--\left( 1-T_S\right)f_-^ev_x+\left( 1-T_D\right)f_+^ev_x + \frac{f_D^e-f_-^e}{\tau_D}L_Cn_{2D} =0
\label{Nmeno}
\end{eqnarray} 
where:
$ f_{D,S}^e=\mathfrak{F}\left(E_+-q\phi_{ch}-\mu_{D,S} \right)$ is the occupation factor at the drain (D) and source (S) contacts, and $\mathfrak{F}$ is Fermi-Dirac distribution function.
\\Let us focus on eq.~\ref{Npiu} (similar considerations can be made for eq.~\ref{Nmeno}):
$J_S^+=T_Sf_S^ev_x$ is the tunneling current component injected from source, $J_D^+=T_Df_+^ev_x$ is instead the drain tunneling current component ejected to the drain, $T_{S,D}$ are the transmission probabilities from source/drain contacts to channel and $v_x$ is the group velocity.
$\left( 1-T_S\right)f_-^ev_x$ and $\left( 1-T_D\right)f_+^ev_x$ are the reflected current components from source and drain barriers, respectively. 
The last term of eqs.~\ref{Npiu} and \ref{Nmeno} is a thermalization process with the source and drain reservoirs, with characteristic times $\tau_S$ and $\tau_D$, respectively.  
The steady-state $f_+^e$  and $f_-^e$ can be obtained by solving eq.~\ref{Npiu} and eq.~\ref{Nmeno}:
\begin{eqnarray}
f_+^{e}=\frac{\left( T_D+\frac{1}{\tau_D\nu}\right)\left(1-T_S \right)f_D^{e}+\left(\frac{1}{\tau_D\nu}+1 \right)\left( T_S+\frac{1}{\tau_S\nu}\right)f_S^{e}}{\left(\frac{1}{\tau_S\nu}+1 \right) \left(\frac{1}{\tau_D\nu}+1 \right) -\left(1-T_S \right)\left(1-T_D \right)},
\end{eqnarray} 
\begin{eqnarray}
f_-^{e}=\frac{\left( T_S+\frac{1}{\tau_S\nu}\right)\left(1-T_D \right)f_S^{e}+\left(\frac{1}{\tau_S\nu}+1 \right)\left( T_D+\frac{1}{\tau_D\nu}\right)f_D^{e}}{\left(\frac{1}{\tau_S\nu}+1 \right) \left(\frac{1}{\tau_D\nu}+1 \right)    -\left(1-T_S \right)\left(1-T_D \right)},
\end{eqnarray} 
where  $\nu=\frac{2\pi^2v_x}{L_C}$ is the inverse of the crossing time $\tau_t$.
The same reasoning can be applied to derive the hole occupation factors in the channel $f_{\pm}^h$.

%
The charge, to be self-consistently solved  with eq.\ref{eq:electrostatics} in order to obtain 
the channel potential $\phi_{ch}$, is computed through the integration on the BZ
\begin{eqnarray}
 Q&=&-\frac{q}{4\pi^2}\int\!\!\!\int_{BZ}{\left(f_+^e+f_-^e \right)  dk_xdk_y }+ \nonumber \\
 &&+\frac{q}{4\pi^2}\int\!\!\!\int_{BZ}{\left(f_+^h+f_-^h \right) dk_xdk_y },
\label{carica_bolla}
\end{eqnarray}
where the total current density is expressed as~\cite{buttiker1988}
\begin{eqnarray}
  J_{tot}&=&\frac{q}{4\pi^2}\left\{\int\!\!\!\int_{BZ} v_x \left(f_+^e-f_-^e \right) dk_x dk_y\right.\nonumber \\
  && \left. +\int\!\!\!\int_{BZ} v_x \left(f_+^h-f_-^h \right) dk_x dk_y \right\}
  \label{corrente}
\end{eqnarray}
%
%
The transmission probability $T_S$ ($T_{D}$) of the interband
barrier at source (drain)  is zero in the source (drain) band gap and
$1$ when there is no barrier between source (drain) and channel. 
When a barrier is present $T_S$ is computed analytically with the WKB
approximation,
assuming $k_y$ conservation due translational invariance along the $y$ direction:
\begin{equation}
  T_S(E, k_y)=\exp{\left\{-2\int_{x_1}^{x_2}|Im(k_x^{E,k_y}(x))|dx\right\}},
  \label{coeff_trasmissione}
\end{equation} 
where $x_1$ and $x_2$ are the classical turning points, and $E$ is the particle kinetic energy.
The same approach is repeated for $T_D$.

The potential profile between each contact and the central region of
the channel is described by an exponential, with
characteristic variation length $\lambda$, obtained
from evanescent mode analysis~\cite{Monroe1998}.
Assuming $t_{sub}\gg t_{ox}> t_{ch}$ we obtain:
\begin{equation}
  \lambda \approx \left(t_{ox}+\frac{t_{ch}}{2}\right)\frac{2}{\pi},
  \label{eq:lambda}
\end{equation}
where $t_{ch}$ is the effective separation between the interfaces of the SiO$_2$ and SiC
layers, for which we assume $t_{ch}=1$~nm~\cite{Dai2008}.

\section{ Electrostatics}  
From analysis of the electrostatics we can gain a better insight of the
device performance limitations.
In fact gate voltage control upon the channel potential (of
which the subthreshold slope $S$ is a measure) is strictly limited by the quantum capacitance $C_q$ of the channel. 
\begin{figure}[!ht]
\centering
\includegraphics[height=5cm]{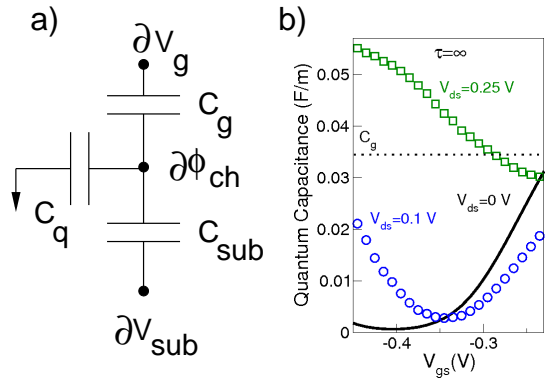}
\caption{a) Equivalent circuit of device electrostatics. b) Quantum capacitance-voltage characteristics for $V_{ds}=0, 0.1, 0.25$~V, $t_{ox}=1$~nm, $t_{sub}=100$~nm and $\alpha_D=9.3\times10^{-3}$ in case of fully ballistic transport. }
\label{capacita}
\normalsize
\end{figure}
Device electrostatics can be schematized as in~\ref{capacita}(a). 
The differential capacitance seen by the gate is 
\begin{equation}
  C_{tg}=C_g\left(1-\frac{\partial \phi_{ch}}{\partial V_{g}} \right)
  \label{Cap_gate_1}
\end{equation}
but, from \ref{capacita}(a), $C_{tg}$  can also be expressed in terms of  capacitances $C_g$,
$C_{sub}$, and $C_q$: 
\begin{equation}
  C_{tg}=\frac{C_g\left( C_{sub}+C_q\right)}{C_g+C_{sub}+C_q}.
  \label{Cap_gate_2}
\end{equation}
From eqs. (\ref{Cap_gate_1}) and (\ref{Cap_gate_2}) we get the
derivative of the channel potential with respect to the gate potential
\begin{equation}
  \frac{\partial \phi_{ch}}{\partial V_{g}}=\frac{C_g}{C_g+C_{sub}+C_q}.
  \label{dphisudvg}
\end{equation}
The expression of the sub-threshold slope $S$ then turns out to be
\begin{equation}
  S=\left(1+\frac{C_{sub}+C_q}{ C_g}\right) \frac{kT}{q}\ln(10),
  \label{S}
\end{equation} 
from which it is clear that $S$ is an increasing function of $C_q$, and
therefore a large quantum capacitance severely limits device performance.

\ref{capacita}(b) shows the capacitance-gate voltage characteristics
for $V_{ds}=0$, $0.1$ and $0.25$~V  obtained  by solving the
Schr\"odinger equation self-consistently with Poisson equation.
In the fully ballistic case the quantum capacitance is low for small $V_{ds}$, indicating a good control of the
channel by the gate voltage, but, as soon as  $V_{ds}$ increases,  hole accumulation in the channel occurs and $C_q$ increases, rapidly degrading $S$ (\ref{capacita}(b)). 
In the inelastic case, instead, the hole accumulation process is slightly suppressed by inelastic injection from the source but the effect on the quantum capacitance is practically negligible. We have observed that for $\tau$ larger than $10^{-4}$~ns, the quantum capacitance basically does not change with respect to ~\ref{capacita}; 
on the other hand, for $\tau < 10^{-4}$~ns $C_q$ decreases with respect to the  fully ballistic case, but the inelastic process becomes dominant and eq.~\ref{S} loses validity.

\section{Perspectives for device operation}
\begin{figure}[!ht]
  \vspace{1cm}
  \centering
  \includegraphics{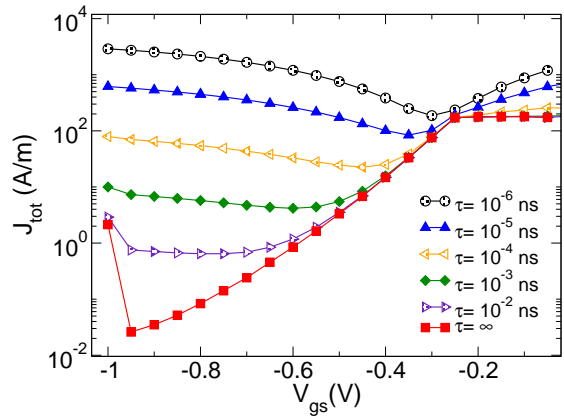}
  \caption{ Transfer-characteristics for $\tau= 10^{-2}, 10^{-3}, 10^{-4}, 10^{-5},10^{-6}$~ns, for $V_{ds}=0.25$~V, $\alpha_D=6.5\times10^{-4}$ and $t_{ox}=1$~nm. }
  \label{bolla}
  \normalsize
\end{figure}

\begin{figure}[!ht]
  \vspace{1cm}
  \centering
  \includegraphics{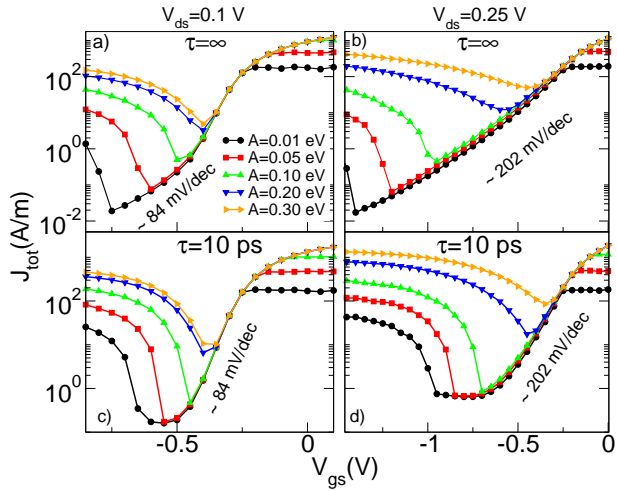}
  \caption{Transfer characteristics for varying with doping parameter $A=0.01$, $0.05$, $0.1$, $0.2$, $0.3$~eV corresponding respectively to $\alpha_D$ of $6.5\times10^{-4}$, $4\times10^{-3}$, $9.3\times10^{-3}$, $2.3\times10^{-2}$, $4.3\times10^{-2}$, $t_{ox}= 2$~nm with: a) $V_{ds}=0.1$~V, $\tau=\infty$, b) $V_{ds}=0.25$~V, $\tau=\infty$, c) $V_{ds}=0.1$~V, $\tau=10$~ps and d) $V_{ds}=0.25$~V, $\tau=10$~ps.}
  \label{Fig4}
  \normalsize
\end{figure}

In order to evaluate the possible performance of the SiC-graphene FET,
we have computed the transfer characteristics by varying three device  
parameters: drain-source voltage 
$V_{ds} = (\mu_S - \mu_D)/q$, donor molar fraction at the contacts $\alpha_D$ (and therefore
parameter $A$) and oxide thickness $t_{ox}$. We also account for
different possible values of inelastic time $\tau$.
First,in \ref{bolla}, we analyze the trend of the transfer characteristics for
different $\tau$ for $V_{ds}=0.25$~V, $t_{ox}=1$~nm and
$\alpha_D=6.5\times10^{-4}$ (corresponding to $A=0.01$~eV).
We observe that for $\tau \ge 1$~ns the transfer characteristics are unaffected and identical to the ballistic case ($\tau\rightarrow \infty$). 
Reducing the relaxation time under $1$~ns, the minimum current increases and the sub-threshold slope remains almost constant  since the quantum capacitance of the channel does not change.
The introduction of inelastic scattering process has mainly two effects in the transfer characteristics: one is a gradual change of the current in the sub-threshold region, the other is an increase of saturation current for $\tau<10^{-4}$~ns or when inelastic current becomes relevant.
In the most favorable case  a sub-threshold slope of $140$~mV/dec can be obtained. 
\\ In \ref{Fig4} we have highlighted the effect of $V_{ds}$ and of the
doping level of contacts.
As expected the main visible effect of increasing $V_{ds}$ is a gradual
degradation of the sub-threshold slope, both in the fully ballistic case
(\ref{Fig4}(a)-(b)) and in the case of  relaxation time $\tau=10$~ps (\ref{Fig4}(c)-(d)), from $84$~mV/dec to $202$~mV/dec.
The reason is simply related to the increased accumulation of holes in the channel with increasing $V_{ds}$, which implies a larger quantum capacitance of the channel and 
therefore a reduced control of the channel potential from the gate voltage.

Increasing the doping causes an increase of both the
maximum current, due to an improved capacity of the source to
inject electrons, and the minimum current.
From \ref{Fig4} we draw the indication that by reducing doping at the contacts we improve the current dynamics.
As already noted, when the source-drain voltage exceeds the gap
of the semiconducting channel ($V_{ds}>0.26$~V), the characteristics
drastically degrade, since band-to-band tunneling current becomes comparable with the thermionic current, 
and hole accumulation in the channel inhibits channel control from the
gate.

\begin{figure}[!h]
  \vspace{1cm}
  \centering
  \includegraphics{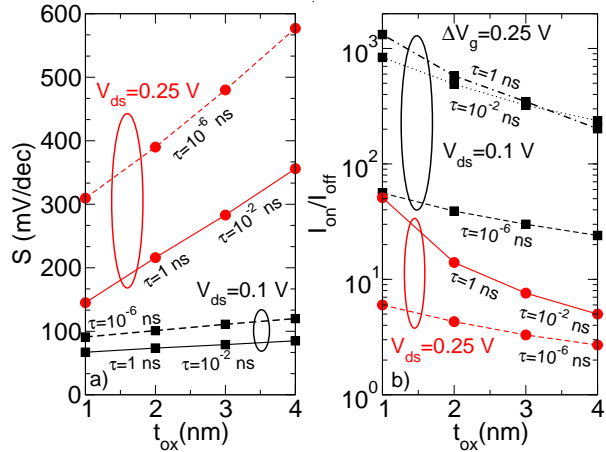}
  \caption{a) Sub-threshold slope for $V_{ds}=0.1$~V and $V_{ds}=0.25$~V calculated with different relaxation time $\tau$. b) $I_{\rm on}/I_{\rm off}$ ratio for $V_{ds}=0.1$~V and $V_{ds}=0.25$~V calculated with different relaxation time $\tau$. }
  \label{Fig5}
  \normalsize
\end{figure}

The increase of oxide thickness $t_{ox}$ has mainly two effects, which can be associated
to a reduction of the capacitive coupling between gate and channel: it increases the
sub-threshold slope $S$ (as shown in  ~\ref{Fig5}(a)), and the opacity of tunneling barriers (i.e. a larger $\lambda$).
The former effect is more evident for $V_{ds}=0.25$~V, where the
quantum capacitance is larger, instead $S$ is almost constant at about $75$~mV/dec for $V_{ds}=0.1$~V and $\tau<10^{-4}$~ns, instead for smaller $\tau$, $S\sim 100$~mV/dec.
~\ref{Fig5}(b) represents the $I_{\rm on}/I_{\rm off}$ ratio as a
function of $t_{ox}$ for $V_{ds}=0.1$~V and $V_{ds}=0.25$~V, calculated
for a gate voltage range $\Delta V_g=0.25$~V for different values of $\tau$.
Larger values of the $I_{\rm on}/I_{\rm off}$ ratio are observed for $\tau=1$~ns.
\\ From our analysis of transfer characteristics, evaluated by varying three device parameters as $V_{ds}$, $\alpha_D$ and $t_{ox}$, we stress the important result that for small  $V_{ds}$ and doping level (\ref{Fig4}(a)) current is modulated by more than 4 orders of magnitude.
\\ We have to stress also the main limitation of graphene on SiC: the energy gap of $0.26$~eV coupled to a low effective mass results in a high band-to-band tunneling current $V_{ds}>0.26$~V and so in an increase of the minimum current achievable.
This limitation on $V_{ds}$ affects the perspectives for  digital circuit operation: in that case we need $V_{ds}=\Delta V_{g}$ and equal to the supply voltage. Even for optimized device parameters ($t_{ox}=1$~nm, $\alpha_D = 4\times10^{-3}$), and a supply voltage of 0.25 V,  we obtain an 
$I_{\rm on}/I_{\rm off}$ ratio of 50, as can been seen from ~\ref{Fig5}(b).

\section{Conclusion}
In this work we have investigated the performance of 
field-effect transistors based on epitaxial graphene on a SiC substrate with an
analytical model.
We have shown that, for small  $V_{ds}$ and doping level, current is modulated by more 
than four orders of magnitude: this is a main improvement with respect to other graphene-based devices~\cite{Lemme2007, Liang2007, Wu2008, Fiori2008}.
Comparable results can be obtained only with carbon nanotubes or graphene nanoribbons, 
but only with post-selection of devices after fabrication (for proper chirality and/or width).
In the case of graphene on SiC, lithography and device patterning are certainly not prohibitive.
A steep subthreshold behavior ($S=67$~mV/decade) can be obtained for
small $V_{ds}=0.1$~V, when the accumulation of holes in the channel is inhibited,
and a larger current ratio, in excess of $10^3$, can be obtained for a gate voltage window of $0.25$~V. 
For digital applications, the limiting factor is represented by the
small voltage drop applicable to the channel, being limited by the
energy gap ($0.26$~eV) of the semiconducting material.
With optimized device parameters we have obtained a sub-threshold slope of $\approx140$~mV/decade and an $I_{\rm on}/I_{\rm off}$ equal to $50$, with a supply voltage of $0.25$~V and $\tau=1$~ns. This falls short of requirements of the International Technology Roadmap for Semiconductors, which requires 
$I_{\rm on}/I_{\rm off} \approx 10^4$~\cite{ITRS2008}.
Finally, we believe that graphene on SiC is very promising as a channel material
for FETs, and much attention has to be put on mechanisms capable to suppress hole injection also at larger
$V_{ds}$, that would allow to improve the subthreshold swing and obtain a good $I_{\rm on}/I_{\rm off}$ also
with a small applied voltage, and on its use in tunnel FETs, where its low gap and low effective mass can be turned into an advantage.

{\bf Acknowledgements}
-The work was supported in part by the the CNR and the EC Sixth Framework Programme, under Contract N. ERAS-CT-2003-980409Õ, through the ESF EUROCORES Programme FoNE-DEWINT, and by the EC Seventh Framework Program under project GRAND (Contract 215752) and the Network of Excellence NANOSIL (Contract 216171).

\clearpage

\bibliography{bibliography_sic}

\end{document}